\documentclass[11pt]{article}

\usepackage{jheppub}
\usepackage{wasysym}

\usepackage{tabularx}
\usepackage{physics}
\usepackage{tikz}

\usepackage{graphicx} 
\usepackage{caption}
\usepackage{subcaption}

\usepackage{ae}

\usepackage{soul}

\usepackage{tabularx}

\usepackage{graphicx}

\usepackage{amsfonts,amsmath,amsthm,amssymb,amsbsy}

\usepackage{mathtools}

\usetikzlibrary{calc}

\newcommand{\dlog}{\dd \hspace{-0.07cm}\log}

\newcommand{\Li}{\text{Li}}

\title{Cosmology meets cluster algebra}
\author[1]{Mattia Capuano,}\emailAdd{m.capuano@herts.ac.uk}
\author[1]{Livia Ferro,}\emailAdd{l.ferro@herts.ac.uk}
\author[1]{Tomasz \L ukowski,}\emailAdd{t.lukowski@herts.ac.uk}
\author[1]{Alessandro Palazio}\emailAdd{a.palazio@herts.ac.uk}

\affiliation[1]{Department of Physics, Astronomy and Mathematics,
University of Hertfordshire,
Hatfield, Hertfordshire, AL10 9AB, United Kingdom}

\abstract{
In this paper we explore the mathematical properties of wavefunction coefficients in power-law FRW cosmologies, and establish their relation to cluster algebras. We focus on the particular contributions to the wavefunction coefficient coming from path graphs, and show that the singularities of the wavefunction associated with an $n$-site path graph are related to the $\mathcal{X}$-coordinates of the cluster algebra $A_{2n-2}$. To establish this relation, we consider the symbol of the de Sitter wavefunction coefficients and show that the letters appearing there are the region variables associated to tubings on the path graph. 
These variables can be rewritten as simplicial coordinates of the moduli space $\mathcal{M}_{0,2n+1}$ and therefore identified with the $\mathcal{X}$-coordinates of type-$A_{2n-2}$ cluster algebras.
We use this result to compute the wavefunction coefficients in terms of cluster functions.}

\begin{document}
\maketitle


\section{Introduction}

Cosmological correlators play a central role in modern cosmology as they encode the properties of the early universe. The standard methods to compute them are usually very involved, which encourages a deeper study of their mathematical structures, in search of novel techniques. For this purpose, a lot of attention has been given to correlators in a simplified cosmological model: a conformally coupled scalar field  in power-law Friedmann-Robertson-Walker (FRW) cosmologies with polynomial interactions.  
This approach follows the same path that has been taken in the study of scattering amplitudes in quantum field theories, where the focus on the toy model of  $\mathcal{N}=4$ super Yang-Mills has allowed for new, interesting mathematical structures to emerge. 
Recently, many novel results in cosmology have been found by transferring methods originating from the scattering amplitude problem in $\mathcal{N}=4$ super Yang-Mills to the computation of cosmological correlators in power-law FRW cosmologies. These include the derivation of differential equations satisfied by the FRW wavefunction coefficients \cite{De:2023xue,De:2024zic,Arkani-Hamed:2023kig,Arkani-Hamed:2023bsv,Baumann:2024mvm,Baumann:2025qjx,Fan:2024iek,Grimm:2024mbw,He:2024olr,Benincasa:2024ptf,Hang:2024xas,Fevola:2024nzj,Capuano:2025ehm}, and positive geometries whose canonical forms compute the integrand of wavefunctions, such as the cosmological polytope for a single wavefunction  \cite{Arkani-Hamed:2017fdk,Arkani-Hamed:2018bjr,Benincasa:2024leu}  and cosmohedra \cite{Arkani-Hamed:2024jbp,Glew:2025otn}.

Another powerful tool that has been successfully used in the scattering amplitudes realm is their relation to cluster algebras, which led to the cluster bootstrap approach to the amplitudes. 
 Cluster algebras \cite{Fomin:2001mwn,williams2013clusteralgebrasintroduction,Fomin:2016caz} are commutative algebras generated by cluster variables, which are Laurent polynomials grouped into sets called clusters. The clusters are related to each other via an operation called mutation, and the full cluster algebra can be computed recursively by starting from a single cluster, the seed, and applying mutations.
Cluster algebras naturally appear in planar $\mathcal{N}=4$ super Yang-Mills, where the locations of singularities of scattering amplitudes are related to cluster coordinates of Grassmannians $\text{Gr}(4,n)$. This addresses the question of which alphabet, i.e. set of letters, appears in the symbol of amplitudes \cite{Golden:2013xva}. The emergence of cluster algebras is significant as they constrain not only which poles and branch cuts appear, but also the order in which they can occur. 
In particular, the notion of cluster adjacency states that cluster variables can appear consecutively
in the symbol of loop amplitudes only if there exists a cluster where they both appear \cite{Drummond:2017ssj}, generalising the Steinmann relations  \cite{Steinmann:1960soa,Steinmann:1960sob} and  constraining consecutive discontinuities of amplitudes.
Cluster adjacency  also appears for tree-level amplitudes, where it controls which  poles can appear in individual BCFW terms \cite{Drummond:2018dfd,Mago:2019waa,Even-Zohar:2023del}.
Therefore, a cluster algebra structure is a powerful tool in constraining the singularities of an amplitude.
The natural question arises whether cluster algebras can also play a central role in the computation of the cosmological observables.

In this paper we uncover the cluster algebra structure behind cosmological correlators in power-law FRW cosmologies. 
The quantity we study is the wavefunction of the universe, whose building blocks are the wavefunction coefficients, and therefore can be expressed in terms of Feynman graphs.
Based on \cite{Capuano:2025ehm}, we start by deriving differential equations for the coefficients in the de Sitter cosmology. This allows us to find the symbol of the de Sitter wavefunction coefficients associated to any tree or loop graph. Focusing on path graphs $P_n$, we show that the symbol can be expressed in terms of the simplicial coordinates on the moduli space of genus-0 curves with $(2n+1)$ marked points. These are known to be associated to $\mathcal{X}$-coordinates of type-$A$ cluster algebras.
This allows us to show that wavefunction coefficients for path graphs $P_n$ with $n$ vertices are related to $A_{2n-2}$ cluster algebras.

This paper is organized as follows. In the next section we set up the background and present a review of the main concepts which will be needed in the rest of the paper. In section \ref{sec:dSsymbols} we derive differential equations in the de Sitter case and find the symbol of the wavefunction coefficients. In section \ref{sec:moduli} we review the relation between the $\mathcal{X}$-coordinates of type-$A$ cluster algebras and the simplicial coordinates on the moduli space of genus-0 curves with $n+3$ marked points. These will be relevant in the following section \ref{sec:cluster}, where we show that the symbol of the wavefunction coefficient associated to the path graph $P_n$ can be expressed in terms of the simplicial coordinates of $\mathcal{M}_{0,2n+1}$. This map automatically establishes a relation between the symbol for the de Sitter wavefunctions and the cluster algebra $A_{2n-2}$.
 In section \ref{sec:applications} we discuss how simplicial coordinates provide a minimal generating set of polylogarithms for the functions to which the symbols integrate, allowing us to perform explicitly the integration. We end with  conclusions and open questions in section \ref{sec:concls}.
 
\paragraph{Note.}
While preparing this draft we have learned that similar results are provided in \cite{XuMazloumi2025}.


\section{Cosmological Background}

Throughout this paper, we will work with a particular class of cosmological toy models. Our main focus will be on a theory of a conformally coupled scalar field $\phi$ with non-conformal polynomial self-interactions in an FRW cosmology in four space-time dimensions, with the action
\begin{equation}
\label{eq:action}
S = - \int d^4x \sqrt{-g} \left(\frac{1}{2} (\partial_\mu \phi)^2+\frac{1}{12}R \phi^2+\sum_{k>2}\frac{\lambda_k}{k!}\phi^k \right),
\end{equation}
where $R$ is the Ricci scalar.
We are interested in correlation functions in power-law cosmologies, evolving as a power law in conformal time $\eta$, with scale factor $a(\eta)=\left(\frac{\eta}{\eta_0}\right)^{-(1+\epsilon)}$. 
Different cosmologies correspond to setting the cosmological parameter $\epsilon$  to different values, with the de Sitter space for $\epsilon=0$, the flat space for $\epsilon=-1$, or an inflationary universe for $\epsilon \sim 0$.   
In these models, correlation functions at a fixed conformal time $\eta=\eta_*=0$ are calculated as functional integrals in terms of the so-called wavefunction of the universe $\Psi[\varphi]$
\begin{equation}
\langle \varphi(\mathbf{x}_1)\ldots  \varphi(\mathbf{x}_N)\rangle=\int \mathcal{D}\varphi\,\varphi(\mathbf{x}_1)\ldots  \varphi(\mathbf{x}_N) |\Psi[\varphi]|^2\,,
\end{equation}
where $\varphi(\mathbf{x})$ is the boundary value of the field $\phi(\eta,\mathbf{x})$, $\varphi(\mathbf{x}) \equiv \phi(0,\mathbf{x})$. Importantly, the wavefunction $\Psi[\varphi]$ admits an expansion in the Fourier space of the form
\begin{equation}
\Psi[\varphi]=\exp\left[-\sum_{n=2}^\infty \int \left(\prod_{a=1}^n\frac{\dd^3 k_a}{(2\pi)^3}\varphi_{\mathbf{k}_a}\right)\psi_n(\mathbf{k}_1,\ldots,\mathbf{k}_n) (2\pi)^3 \delta^3(\mathbf{k}_1+\ldots +\mathbf{k}_n)\right] \,,
\end{equation}
where the three-vectors ${\bf{k}}_a$ are the spatial momenta, whose magnitudes $|{\bf{k}}_a|$ are referred to as ``energies''.
 The functions $\psi_n(\mathbf{k}_1,\ldots,\mathbf{k}_n)$, called wavefunction coefficients,  can be represented as sums of Feynman diagrams and computed using Feynman rules, similar to those used for scattering amplitudes in quantum field theories. 
Importantly, the function associated to a Feynman graph $G$  depends only on the variables $X_v$, i.e. the sums of the external energies entering each vertex $v$, and on $Y_e$, i.e. the energies associated to each internal edge $e$ of the graph. 
Finally, the wavefunction coefficient for a given graph $G$ for generic FRW cosmologies can be constructed from its flat-space counterpart, $\epsilon=-1$, 
by shifting the external energies $X_v\rightarrow X_v+x_v$ and performing the integral over $x_v$ with the twist function $u=\prod_{v} x_v^{\alpha_v}$:
\begin{equation}\label{eq:FRWintro}
\psi_G^{\text{FRW}}(X_v,Y_e)=\int_0^\infty \left(\prod_v \dd x_v \,x_v^{\alpha_v}\right) \psi_G^{\text{flat}} (X_v+x_v,Y_e) \,,
\end{equation}
where the parameters ${\alpha_v}$ are related to the cosmological parameter $\epsilon$ and the valency of the vertices. 
Interestingly, the flat-space wavefunction coefficients have a geometric description in terms of the cosmological polytopes \cite{Arkani-Hamed:2017fdk}.


\section{Symbols for de Sitter}
\label{sec:dSsymbols}

Our main goal is to study the singularity structure of the general cosmological wavefunctions \eqref{eq:FRWintro}. Importantly, these singularities are captured by the (linear) differential equations satisfied by $\psi_G^{\text{FRW}}$ \cite{Hillman:2019wgh,Arkani-Hamed:2023kig,Capuano:2025ehm}, and are related to the linear coefficients in these equations. To focus our attention, in this section we only consider the de Sitter space wavefunctions in a theory with cubic interaction, for which $\epsilon=0$ and all $\alpha_v=0$:
\begin{equation}\label{eq:dS1}
\psi_G^{\text{dS}}(X_v,Y_e)=\int_0^\infty \prod_v \dd x_v\, \psi_G^{\text{flat}} (X_v+x_v,Y_e)=:\int_0^\infty  \Omega_G^{\text{flat}} (X_v+x_v,Y_e) \,.
\end{equation}
This will be sufficient to understand the general structure of singularities since the coefficients appearing for the de Sitter case provide a complete set of functions appearing in the generic case. 

In particular, in this section we propose a modification of the method that we introduced in \cite{Capuano:2025ehm} for generic FRW cosmologies, to derive the differential equations in the de Sitter case. This modification is necessary since the case when $\epsilon=0$ is not generic, and the arguments concerning boundary conditions that we have used in \cite{Capuano:2025ehm} do not hold when $\epsilon=0$. Additionally, we will find the symbol of the de Sitter wavefunction coefficients associated to any tree or loop graph. Importantly, our results agree with those previously found in the literature \cite{Hillman:2019wgh, Arkani-Hamed:2023kig, He:2024olr}.

Our construction and definitions will closely follow the ones in \cite{Capuano:2025ehm}. In particular, we will use the notion of graph tubes and graph tubings, as explained in the following. Let $G$ be any graph. We denote by $V_G$ the vertex set of $G$ and by $E_G$ the edge set of $G$. A tube $T$ on $G$ is a connected subgraph induced by a subset of vertices of $G$. The tube induced by the whole graph $G$ is called the \textit{root}. A $u$-tubing $\tau$ on $G$ is a collection of compatible tubes $\{T_i\}_{i=1,\ldots,p}$, where two tubes $T_i$ and $T_j$ are compatible if either one is included in another (i.e. $T_i\subset T_j$ or $T_j\subset T_i$), or they are not adjacent (i.e. the vertex set of $T_i\cup T_j$ does not induce a connected subgraph of $G$). A maximal set of compatible tubes is called a \textit{maximal} $u$-tubing. We denote the set of all maximal $u$-tubings as $\mathcal{U}_G^{\text{max}}$. 

To each vertex $v\in V_G$ we associate a variable $X_v$, and to each edge $e\in E_G$ we associate a variable $Y_e$. Then, to each tube $T$, we can associate the linear function
\begin{equation}\label{eq:HT}
H_T=\sum_{v\in V_T}\left(X_v+\sum_{e=[v,v'] \in G\setminus T} Y_e\right),
\end{equation}
where the second sum runs over the edges incident to $v$ in $G\setminus T$, and self-loops are counted twice.

Additionally, for any tubing $\tau$, there is a natural map
\begin{equation}\label{eq:vertexbijection}
q_{\tau}: V_G \to \tau\,,
\end{equation}
that associates to $v\in V_G$ the smallest (by inclusion) tube $T_v\in\tau$ that contains $v$.
Given a tubing $\tau$ and a vertex $v$, we define a \textit{region} $r_{v,\tau}$ to be the subset of the tubing, $r_{v,\tau}=\{T_1;T_2,\ldots , T_{|r_{v,\tau}|}\}\subset\tau$, such that $T_1=T_v$, $T_j\subset T_v$, and the $T_j$'s are the subset of $\tau$ of the largest (by inclusion) disjoint tubes. We call $T_v$ the \textit{parent tube}, and all other $T_j$'s the \textit{children tubes}.
To each region $r_{v,\tau}$ we associate the linear function $R_{v,\tau}$ using the $H_T$ functions defined in \eqref{eq:HT}
\begin{equation}
R_{v,\tau}=H_{T_v}-\sum_{j=2}^{|r|}H_{T_{j}}\,,
\end{equation}
which subtracts all children tube variables from the parent tube variable.
We refer to $R_{v,\tau}$ as \textit{region variables}. The region variables will play a crucial role in our construction, and they will provide a complete set of singularities of the wavefunction coefficients associated to a graph $G$.

In order to derive differential equations for the de Sitter wavefunction $\psi_G^{\text{dS}}$, we will use its decomposition into amplitubes \cite{Glew:2025ugf}. First, for a given tubing $\tau$, we define a logarithmic form
\begin{equation}
    \mu_{\tau}=\bigwedge_{T\in\tau}\dlog {H_T}\,,
\end{equation}
where the order in the wedge product is determined by a fixed ordering of vertices of $G$, together with the function $q_\tau$ in \eqref{eq:vertexbijection}. The differential $\text{d}$ is defined to act only on the vertex variables $X_v$. Additionally, for a given graph $G$, we define the \textit{amplitube}
\begin{equation}
\omega_G=\sum_{\tau\in\mathcal{U}_G^{\text{max}}}\mu_{\tau}\,,
\end{equation}
where the sum is over all maximal $u$-tubings on $G$.
 Then, the flat-space wavefunction coefficient associated to the graph $G$ is the combination of the amplitubes of the subgraphs of $G$ obtained from it by cutting all possible subsets of edges
\begin{equation}\label{eq: big Omega}
    \Omega_G^{\text{flat}}(X_v,Y_e)=\sum_{\mathbf{e}\subseteq E_G}(-1)^{|\mathbf{e}|}\sum_{\tau\in\mathcal{U}_{G_{\mathbf{e}}}^{\text{max}}} \mu_{\tau}=\psi_G^{\text{flat}}(X_v,Y_e)\text{d}X_1\wedge\dots\wedge\text{d}X_{|V_G|}\,,
\end{equation}
where $G_\mathbf{e}=G[E_G\setminus \mathbf{e}]$.
The wavefunction coefficient for de Sitter is then
\begin{equation}
    \psi_G^{\text{dS}}(X_v,Y_e)=\int_{x_v\geq0} \prod_{i=1}^{|V_G|}\text{d}x_i\,\psi^{\text{flat}}_G(X_v+x_v,Y_e)=\int_{x_v\geq0}\Omega_G^{\text{flat}}(X_v+x_v,Y_e),    
\end{equation}
where now we take the differential d to act on $x_i$'s. 

Applying a similar method to the one we used in \cite{Capuano:2025ehm}, we can derive a set of differential equations satisfied by the de Sitter wavefunction coefficients. Then, straightforwardly, we will be able to find the symbol associated to these coefficients, which has previously been found in \cite{Hillman:2019wgh}, \citep{Arkani-Hamed:2023kig} and \cite{He:2024olr}. To derive these equations, we start by acting with the kinematic differential $\text{d}_{\text{kin}}=\sum \frac{\partial}{\partial X_v}\text{d}X_v+\frac{\partial}{\partial Y_e}\text{d}Y_e$ on an integral of the form 
\begin{equation}
\psi(X_v,Y_e)=\int_{x_v\geq 0} \Omega(x_v+X_v,Y_e)\,,
\end{equation}
where $\Omega$ is a logarithmic differential form in $x_v$. Then, by direct calculations and using the fact that $\Omega$ is logarithmic, we arrive at
\begin{equation}
    \text{d}_{\text{kin}}\psi(X_v,Y_e)=\int_{x_v\geq0}\text{d}_{\text{kin}}\,\Omega(x_v+X_v,Y_e)=-\int_{x_v\geq0}\text{d}\,\tilde{\Omega}(x_v+X_v,Y_e)\,,
\end{equation}
where $\tilde{\Omega}=\Omega\big|_{\text{d}\to\mathcal{D}}$ and $\mathcal{D}=\text{d}+\text{d}_{\text{kin}}$.
Finally, with integration by parts we find
\begin{equation}\label{eq:int_by_parts}
    \text{d}_{\text{kin}}\psi(X_v,Y_e)=-\sum_{v}(-1)^{|v|}\int_{x_v\geq0}\tilde{\Omega}(x_v+X_v,Y_e)\,\delta(x_v)\,,
\end{equation}
where we introduced an order on the vertex set $V_G$ and $|v|$ indicates the position of the vertex $v$ in this order. We have dropped in \eqref{eq:int_by_parts} the boundary terms at infinity since they cancel for the physical combinations. Importantly, the right-hand side of \eqref{eq:int_by_parts} contains lower dimensional integrals, since one of the $x_v$ variables is localised at 0. When this method is applied to integrals of logarithmic forms $\mu_\tau$, the right hand side can be again expanded in terms of integrals of forms associated to $u$-tubings, with one tube in $\tau$ removed. Using this approach, we have been able to derive the following explicit set of differential equations for a given $F_{\tau}=\int \mu_\tau$ 
\begin{equation}\label{eq: master}
    \text{d}_{\text{kin}}F_{\tau}=\sum_{T\in \tau}\dd_{\text{kin}}\text{log}(R_{T,\tau})\sum_{T'\in r_{T,\tau}} \text{gen}(T') F_{\tau\setminus T'}\,,
\end{equation}
where $R_{T,\tau}$ is the region variable of the region $r_{T,\tau}$ whose parent tube is $T$, $\tau\setminus T'$ is the tubing obtained from $\tau$ by removing the tube $T'$, and $\text{gen}(T')$ is $+1$ if $T'$ is the parent tube of $r_{T,\tau}$ and $-1$ otherwise.

For example, consider the path graph with two vertices. Then the de Sitter wavefunction coefficient associated to this graph can be expanded as
\begin{equation}
\psi_{\begin{tikzpicture}[scale=0.6]
    \coordinate (A) at (0,0);
    \coordinate (B) at (1/2,0);
    \draw[thick] (A) -- (B);
    \fill[black] (A) circle (2pt);
    \fill[black] (B) circle (2pt);
\end{tikzpicture}}^{\mathrm{dS}}=F_{\begin{tikzpicture}[scale=0.6]
    \coordinate (A1) at (0,0);
    \coordinate (B1) at (1,0);
    \draw[thick] (A1) -- (B1);
        \draw[thick, black] (1/2,0) ellipse (0.75cm and 0.3cm);
            \draw[thick, black] (0,0) ellipse (0.15cm and 0.15cm);
    \fill[black] (A1) circle (2pt);
    \fill[black] (B1) circle (2pt);
\end{tikzpicture}}+F_{\begin{tikzpicture}[scale=0.6]
    \coordinate (A1) at (0,0);
    \coordinate (B1) at (1,0);
    \draw[thick] (A1) -- (B1);
        \draw[thick, black] (1/2,0) ellipse (0.75cm and 0.3cm);
            \draw[thick, black] (1,0) ellipse (0.15cm and 0.15cm);
    \fill[black] (A1) circle (2pt);
    \fill[black] (B1) circle (2pt);
\end{tikzpicture}}-F_{\begin{tikzpicture}[scale=0.6]
    \coordinate (A1) at (0,0);
    \coordinate (B1) at (1,0);
    \draw[thick] (A1) -- (B1);
            \draw[thick, black] (0,0) ellipse (0.15cm and 0.15cm);
            \draw[thick, black] (1,0) ellipse (0.15cm and 0.15cm);
    \fill[black] (A1) circle (2pt);
    \fill[black] (B1) circle (2pt);
\end{tikzpicture}}\,,
\end{equation}
where we have introduced the following set of integrals
\begin{align}
    F_{\begin{tikzpicture}[scale=0.6]
    \coordinate (A1) at (0,0);
    \coordinate (B1) at (1,0);
    \draw[thick] (A1) -- (B1);
        \draw[thick, black] (1/2,0) ellipse (0.75cm and 0.3cm);
            \draw[thick, black] (0,0) ellipse (0.15cm and 0.15cm);
    \fill[black] (A1) circle (2pt);
    \fill[black] (B1) circle (2pt);
\end{tikzpicture}}&=\int_{x_1\geq0}\int_{x_2\geq0} \dd\log(x_1+X_1+Y_{1,2})\wedge\dd\log(x_1+x_2+X_1+X_2)\,,\\
        F_{\begin{tikzpicture}[scale=0.6]
    \coordinate (A1) at (0,0);
    \coordinate (B1) at (1,0);
    \draw[thick] (A1) -- (B1);
        \draw[thick, black] (1/2,0) ellipse (0.75cm and 0.3cm);
            \draw[thick, black] (1,0) ellipse (0.15cm and 0.15cm);
    \fill[black] (A1) circle (2pt);
    \fill[black] (B1) circle (2pt);
\end{tikzpicture}}&=\int_{x_1\geq0}\int_{x_2\geq0} \dd\log(x_1+x_2+X_1+X_2)\wedge\dd\log(x_2+X_2+Y_{1,2})\,,\\
            F_{\begin{tikzpicture}[scale=0.6]
    \coordinate (A1) at (0,0);
    \coordinate (B1) at (1,0);
    \draw[thick] (A1) -- (B1);
            \draw[thick, black] (0,0) ellipse (0.15cm and 0.15cm);
            \draw[thick, black] (1,0) ellipse (0.15cm and 0.15cm);
    \fill[black] (A1) circle (2pt);
    \fill[black] (B1) circle (2pt);
\end{tikzpicture}}&=\int_{x_1\geq0}\int_{x_2\geq0} \dd\log(x_1+X_1+Y_{1,2})\wedge\dd\log(x_2+X_2+Y_{1,2})\,.
\end{align}
Using \eqref{eq: master},
\begin{align}   &\dd_{\text{kin}}F_{\begin{tikzpicture}[scale=0.6]
    \coordinate (A1) at (0,0);
    \coordinate (B1) at (1,0);
    \draw[thick] (A1) -- (B1);
        \draw[thick, black] (1/2,0) ellipse (0.75cm and 0.3cm);
            \draw[thick, black] (0,0) ellipse (0.15cm and 0.15cm);
    \fill[black] (A1) circle (2pt);
    \fill[black] (B1) circle (2pt);
\end{tikzpicture}}=\dd_{\text{kin}}\log(X_1+Y_{1,2})F_{\begin{tikzpicture}[scale=0.6]
    \coordinate (A1) at (0,0);
    \coordinate (B1) at (1,0);
    \draw[thick] (A1) -- (B1);
        \draw[thick, black] (1/2,0) ellipse (0.75cm and 0.3cm);
    \fill[black] (A1) circle (2pt);
    \fill[black] (B1) circle (2pt);
\end{tikzpicture}}+\dd_{\text{kin}}\log(X_2-Y_{1,2})(-F_{\begin{tikzpicture}[scale=0.6]
    \coordinate (A1) at (0,0);
    \coordinate (B1) at (1,0);
    \draw[thick] (A1) -- (B1);
        \draw[thick, black] (1/2,0) ellipse (0.75cm and 0.3cm);
    \fill[black] (A1) circle (2pt);
    \fill[black] (B1) circle (2pt);
\end{tikzpicture}}+F_{\begin{tikzpicture}[scale=0.6]
    \coordinate (A2) at (0,0);
    \coordinate (B2) at (1,0);
    \draw[thick] (A2) -- (B2);
    \draw[thick, black] (0,0) ellipse (0.15cm and 0.15cm);
    \fill[black] (A2) circle (2pt);
    \fill[black] (B2) circle (2pt);
\end{tikzpicture}})\,,\\
 &\text{d}_\text{kin}F_{\begin{tikzpicture}[scale=0.6]
    \coordinate (A1) at (0,0);
    \coordinate (B1) at (1,0);
    \draw[thick] (A1) -- (B1);
        \draw[thick, black] (1/2,0) ellipse (0.75cm and 0.3cm);
            \draw[thick, black] (1,0) ellipse (0.15cm and 0.15cm);
    \fill[black] (A1) circle (2pt);
    \fill[black] (B1) circle (2pt);
\end{tikzpicture}}=\dd_{\text{kin}}\log(X_2+Y_{1,2})F_{\begin{tikzpicture}[scale=0.6]
    \coordinate (A1) at (0,0);
    \coordinate (B1) at (1,0);
    \draw[thick] (A1) -- (B1);
        \draw[thick, black] (1/2,0) ellipse (0.75cm and 0.3cm);
    \fill[black] (A1) circle (2pt);
    \fill[black] (B1) circle (2pt);
\end{tikzpicture}}+\dd_{\text{kin}}\log(X_1-Y_{1,2})(-F_{\begin{tikzpicture}[scale=0.6]
    \coordinate (A1) at (0,0);
    \coordinate (B1) at (1,0);
    \draw[thick] (A1) -- (B1);
        \draw[thick, black] (1/2,0) ellipse (0.75cm and 0.3cm);
    \fill[black] (A1) circle (2pt);
    \fill[black] (B1) circle (2pt);
\end{tikzpicture}}+F_{\begin{tikzpicture}[scale=0.6]
    \coordinate (A3) at (4,0);
    \coordinate (B3) at (5,0);
    \draw[thick] (A3) -- (B3);
    \draw[thick, black] (5,0) ellipse (0.15cm and 0.15cm);
    \fill[black] (A3) circle (2pt);
    \fill[black] (B3) circle (2pt);
\end{tikzpicture}})\,,\\
  &\text{d}_\text{kin}F_{\begin{tikzpicture}[scale=0.6]
    \coordinate (A1) at (0,0);
    \coordinate (B1) at (1,0);
    \draw[thick] (A1) -- (B1);
            \draw[thick, black] (0,0) ellipse (0.15cm and 0.15cm);
            \draw[thick, black] (1,0) ellipse (0.15cm and 0.15cm);
    \fill[black] (A1) circle (2pt);
    \fill[black] (B1) circle (2pt);
\end{tikzpicture}}=\dd_{\text{kin}}\log(X_1+Y_{1,2})F_{\begin{tikzpicture}[scale=0.6]
    \coordinate (A3) at (4,0);
    \coordinate (B3) at (5,0);
    \draw[thick] (A3) -- (B3);
    \draw[thick, black] (5,0) ellipse (0.15cm and 0.15cm);
    \fill[black] (A3) circle (2pt);
    \fill[black] (B3) circle (2pt);
\end{tikzpicture}}+\dd_{\text{kin}}\log(X_2+Y_{1,2})F_{\begin{tikzpicture}[scale=0.6]
    \coordinate (A2) at (0,0);
    \coordinate (B2) at (1,0);
    \draw[thick] (A2) -- (B2);
    \draw[thick, black] (0,0) ellipse (0.15cm and 0.15cm);
    \fill[black] (A2) circle (2pt);
    \fill[black] (B2) circle (2pt);
\end{tikzpicture}}\,,
\end{align}
where
\begin{align}
F_{\begin{tikzpicture}[scale=0.6]
    \coordinate (A1) at (0,0);
    \coordinate (B1) at (1,0);
    \draw[thick] (A1) -- (B1);
        \draw[thick, black] (1/2,0) ellipse (0.75cm and 0.3cm);
    \fill[black] (A1) circle (2pt);
    \fill[black] (B1) circle (2pt);
\end{tikzpicture}}&=\int_{x>0}\dd\log(x+X_1+X_2)\,,\\
F_{\begin{tikzpicture}[scale=0.6]
    \coordinate (A2) at (0,0);
    \coordinate (B2) at (1,0);
    \draw[thick] (A2) -- (B2);
    \draw[thick, black] (0,0) ellipse (0.15cm and 0.15cm);
    \fill[black] (A2) circle (2pt);
    \fill[black] (B2) circle (2pt);
\end{tikzpicture}}&=\int_{x>0}\dd\log(x+X_1+Y_{1,2})\,,\\
F_{\begin{tikzpicture}[scale=0.6]
    \coordinate (A3) at (4,0);
    \coordinate (B3) at (5,0);
    \draw[thick] (A3) -- (B3);
    \draw[thick, black] (5,0) ellipse (0.15cm and 0.15cm);
    \fill[black] (A3) circle (2pt);
    \fill[black] (B3) circle (2pt);
\end{tikzpicture}}&=\int_{x>0}\dd\log(x+X_2+Y_{1,2})\,.
\end{align}

By iterating this process, namely by acting with the kinematic differential $\dd_{\text{kin}}$ on the functions appearing on the right-hand side of the previous step, we eventually remove all integrations. In this way, if we introduce a set of integrals $F_\tau=\int \mu_\tau$ for all $u$-tubings of $G_\mathbf{e}$, for all subsets $\mathbf{e}$ of $E_G$, we are able to derive a closed set of differential equations. These equations will be in a general form 
\begin{equation}\label{eq:diff_gen}
    \dd_{\text{kin}} F_\tau=\sum_{\tilde\tau}( \dd_{\text{kin}}\log f_{\tau\tilde{\tau}})F_{\tilde{\tau}}\,,
\end{equation}
where $f_{\tau\tilde{\tau}}$ is one of the region variables of the graph $G$.

Knowing the differential equations \eqref{eq:diff_gen}, the symbol obeys
\begin{equation}\label{eq: symbol recurrence}
    \mathcal{S}(F_{\tau})=\sum_{\tilde{\tau}} \mathcal{S}(F_{\tilde{\tau}})\otimes f_{\tau\tilde{\tau}}\,.
\end{equation}
This property allows us to find the symbols for $\psi^{\text{dS}}_G(X_v,Y_e)$.

For the example of the path graph with two vertices that we studied above, this immediately gives
\begin{equation}
    \mathcal{S}(F_{\begin{tikzpicture}[scale=0.6]
    \coordinate (A1) at (0,0);
    \coordinate (B1) at (1,0);
    \draw[thick] (A1) -- (B1);
        \draw[thick, black] (1/2,0) ellipse (0.75cm and 0.3cm);
            \draw[thick, black] (0,0) ellipse (0.15cm and 0.15cm);
    \fill[black] (A1) circle (2pt);
    \fill[black] (B1) circle (2pt);
\end{tikzpicture}})=(X_1+X_2)\otimes(X_1+Y_{1,2})-(X_1+X_2)\otimes(X_2-Y_{1,2})+(X_1+Y_{1,2})\otimes(X_2-Y_{1,2})\,.
\end{equation}
This method works for all graphs, both at tree and loop level, and we have compared our results with those available in the literature \cite{Hillman:2019wgh,Arkani-Hamed:2023kig} and found that they agree.

The most important outcome of the calculation in this section is the formula \eqref{eq: master}, which shows that every letter in the symbol for any de Sitter wavefunction coefficient is a region variable for one of the regions of the underlying graph. While the calculations in this section can be applied to any graph $G$, in the remaining sections of this paper we will focus our attention on path graphs $P_n$ for $n=1,2,\ldots$, and show how the region variables, and therefore the symbol letters, relate to the type-$A$ cluster algebras.


\section{$A_n$ Cluster algebras and $\mathcal{M}_{0,n+3}$}
\label{sec:moduli}

In this section we briefly review the relation between the $\mathcal{X}$-coordinates of type-$A$ cluster algebras and the simplicial coordinates on the moduli space of genus-0 curves with $n+3$ marked points, $\mathcal{M}_{0,n+3}$. These will be relevant in the following section, where we show that the symbol of the wavefunction coefficient associated to the path graph $P_n$ can be expressed in terms of the simplicial coordinates of $\mathcal{M}_{0,2n+1}$. This map automatically establishes a relation between the symbol for the de Sitter wavefunctions for path graphs and the cluster algebras $A_{2n-2}$. The result extends to power-law FRW cosmologies since the alphabet is the same.
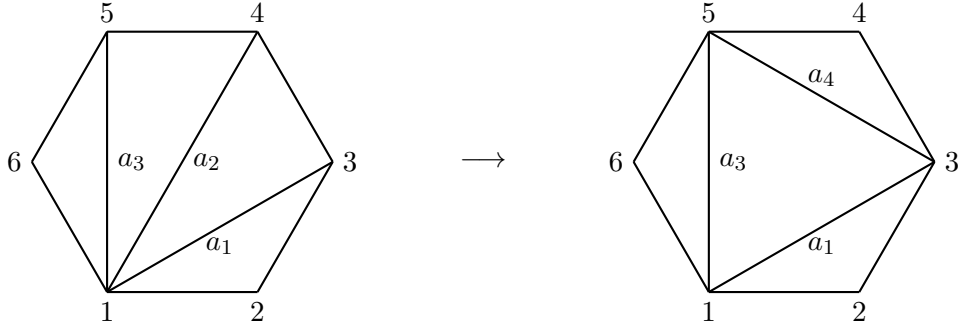
\begin{figure}
    \centering
    \usetikzlibrary {arrows.meta}
\begin{tikzpicture}[scale=2]
    \coordinate (A) at (1,0);
    \coordinate (B) at ({cos(60)},{sin(60)});
    \coordinate (C) at ({cos(120)},{sin(120)});
    \coordinate (D) at ({cos(180)},{sin(180)});
    \coordinate (E) at ({cos(240)},{sin(240)});
    \coordinate (F) at ({cos(300)},{sin(300)});
    \node[right] at (A) {3};
    \node[above] at (B) {4};
    \node[above] at (C) {5};
    \node[left] at (D) {6};
    \node[below] at (E) {1};
    \node[below] at (F) {2};
    \draw[thick] (A) -- (B);
    \draw[thick] (B) -- (C);
    \draw[thick] (C) -- (D);
    \draw[thick] (D) -- (E);
    \draw[thick] (E) -- (F);
    \draw[thick] (F) -- (A);
    
    \draw[thick] (E)--(A) node[midway,below]{$a_1$};
    \draw[thick] (E)--(B) node[midway,right]{$a_2$};
    \draw[thick] (E)--(C) node[midway,right]{$a_3$};

    \node at (2,0) {$\longrightarrow$};

    \coordinate (G) at (1+4,0);
    \coordinate (H) at ({cos(60)+4},{sin(60)});
    \coordinate (I) at ({cos(120)+4},{sin(120)});
    \coordinate (J) at ({cos(180)+4},{sin(180)});
    \coordinate (K) at ({cos(240)+4},{sin(240)});
    \coordinate (L) at ({cos(300)+4},{sin(300)});
    \node[right] at (G) {3};
    \node[above] at (H) {4};
    \node[above] at (I) {5};
    \node[left] at (J) {6};
    \node[below] at (K) {1};
    \node[below] at (L) {2};
    \draw[thick] (G) -- (H);
    \draw[thick] (H) -- (I);
    \draw[thick] (I) -- (J);
    \draw[thick] (J) -- (K);
    \draw[thick] (K) -- (L);
    \draw[thick] (L) -- (G);
    
    \draw[thick] (K)--(G) node[midway,below]{$a_1$};
    \draw[thick] (G)--(I) node[midway,above]{$a_4$};
    \draw[thick] (K)--(I) node[midway,right]{$a_3$};
    
\end{tikzpicture}
    \caption{A flip of the diagonal $(14)$ in the quadrilateral $(1345)$ corresponds to a mutation of the $a_2$ variable in the seed cluster $\{a_1,a_2,a_3\}$ into the cluster \{$a_1,a_4,a_3$\}. The cross-ratios associated to each quadrilateral \eqref{eq: cross ratios} mutate like $\mathcal{X}$-coordinates.   }\label{fig: diagonal flip}
\end{figure}

We will not provide a detailed introduction to cluster algebra, and we refer the reader to established literature on this subject \cite{williams2013clusteralgebrasintroduction}. However, we will recall a few basic facts about type-$A$ cluster algebras, mostly regarding the combinatorics of variables in cluster algebras, which will be relevant later in our discussion. 
First, for type-$A_n$ cluster algebras, each cluster can be conveniently represented as a triangulation of an $(n+3)$-\text{gon}. In this picture, the diagonals in the triangulation correspond to the $\mathcal{A}$-coordinates of that cluster and a cluster mutation is equivalent to a diagonal flip, as shown in figure \ref{fig: diagonal flip}.
In a given triangulation, one can also associate to each diagonal a $\mathcal{X}$-coordinate. The triangulation and the diagonal define a unique quadrilateral for which the diagonal of the polygon is also the diagonal of the quadrilateral. Let $i,j,k,l$ be the vertices of the quadrilateral with ordering $i<j<k<l$, up to cyclic symmetry. We require that $i$ be one of the two endpoints of the diagonal. Then, define the cross-ratio
\begin{equation}\label{eq: cross ratios}
    z_{(i,j;k,l)}=\frac{(z_i-z_j)(z_k-z_l)}{(z_i-z_l)(z_j-z_k)}\,,
\end{equation}
where $z_m$'s are variables associated to vertices of the polygon. Under a geometric flip these cross-ratios mutate as the corresponding $\mathcal{X}$-coordinates of the cluster algebra $A_{n}$. Let us illustrate it using the cluster algebra $A_3$ as an example. In this case, we are interested in a hexagon and its triangulations. Let us take the triangulation of the hexagon with chords $(13), (14)$ and $(15)$, which is often referred to as the seed triangulation. In this case, we have the following cross-ratios 
\begin{align}\label{xToz}
    z_{(1,2;3,4)}=\frac{(z_1-z_2)(z_3-z_4)}{(z_1-z_4)(z_2-z_3)},\,\,\,\, && z_{(1,3;4,5)}=\frac{(z_1-z_3)(z_4-z_5)}{(z_1-z_5)(z_3-z_4)},&&
    z_{(1,4;5,6)}=\frac{(z_1-z_4)(z_5-z_6)}{(z_1-z_6)(z_4-z_5)}.
\end{align}
Under the flip shown in figure \ref{fig: diagonal flip}, $z_{(1,3;4,5)}$ transforms into $z_{(3,4;5,1)}$, with the relation given by 
\begin{equation}
z_{(1,3;4,5)}\to z_{(3,4;5,1)}=\frac{1}{z_{(1,3;4,5)}} \,.
\end{equation}
This is what we expect from the $\mathcal{X}$-coordinates mutation rules.

Let us now establish the connection between type-$A$ cluster algebras and moduli spaces of genus-0 curves with marked points. First, we define the simplicial coordinates on $\mathcal{M}_{0,{n+3}}$, that is the configuration space of $n+3$ points on the complex projective line $\mathbb{P}_{\mathbb{C}}^1$ modulo the action of the projective linear group $\text{PGL}{(2,\mathbb{C})}$, that is, the group of automorphisms of $\mathbb{P}_{\mathbb{C}}^1$: 
\begin{equation}
    \mathcal{M}_{0,n+3}:=\{(z_1,\dots,z_{n+3})\in (\mathbb{P}_{\mathbb{C}}^1)^{n+3}, z_i\neq z_j \text{ for }i\neq j\}/\text{PGL}{(2,\mathbb{C})}\,.
\end{equation}
Since functions in the moduli space must be invariant under $\text{PGL}{(2,\mathbb{C})}$, they must depend only on combinations of cross-ratios that are invariant under such transformations. Then, we see that equations such as \eqref{eq: cross ratios} define a map between the $\mathcal{X}$-coordinates of $A_n$ and the cross-ratios of $\mathcal{M}_{0,n+3}$. This correspondence has been generalized to other cluster algebras in \cite{Arkani-Hamed:2020tuz}.

We can go further, and map each seed of the cluster algebra to a coordinate of the moduli space. First, the action of the automorphism group can be used to fix three points to 0, 1 and $\infty$, such that
\begin{equation}
    \mathcal{M}_{0,n+3}=\{(z_1,\dots,z_{n})\in\mathbb{C}^{n}:z_i\neq z_j, z_i\neq 0,1\}\,,
\end{equation}
and the space can be regarded as the complement in $\mathbb{C}^n$ of an arrangement of affine hyperplanes 
\begin{equation}\label{eq: hyperplanes}
\mathcal{M}_{0,n+3}=\mathbb{C}^{n}\backslash\{z_i=0,z_i=1,z_i=z_j\}\,,
\end{equation}
with $i,j=1,\dots,n$. The coordinates $\{z_1,\dots,z_n\}$ are called \textit{simplicial coordinates} because in the real moduli space $\mathcal{M}_{0,n+3}(\mathbb{R})$, the arrangement of hyperplanes defines a simplex.
By applying the same gauge fixing on equations such as \eqref{xToz}, the $\mathcal{X}$-coordinates can be expressed in terms of linear forms defining the hyperplanes \eqref{eq: hyperplanes}. 
Finally, once a gauge fixing has been applied, relations of the type \eqref{xToz} can be inverted, providing a way to map each simplicial coordinate into a combination of $\mathcal{X}$-coordinates. This also provides an efficient tool to compute $\mathcal{X}$-coordinates as a function of the seed variables without resorting to any mutation: it is sufficient to list all the cross-ratios and, after choosing a gauge, use the inverse of \eqref{xToz} to express each cross-ratio in the seed variables.

A different way to map $\mathcal{X}$-coordinates into simplicial coordinates using the \textit{hedgehog} construction was shown in \cite{Parker:2015cia}. Simplicial coordinates have also been applied to study the cluster structures of single Feynman integrals in \cite{Chicherin:2020umh}.


\section{Cluster algebras for path graphs}
\label{sec:cluster}
In section \ref{sec:dSsymbols}, we have found that all letters appearing in the symbol of the de Sitter wavefunction coefficients $\psi_{G}^{\text{dS}}$ are region variables of the graph $G$. Importantly, in the symbol for these functions not all region variables are present in the final answer. In particular, those that are present in the final answer are consistent with those found in \cite{Hillman:2019wgh} as well as in the kinematic flow approach \cite{Arkani-Hamed:2023kig}. 

In this section we will focus on a subset of possible graphs: the path graphs $P_n$. We order the vertices in $P_n$ from left to right, associate the variable $X_i$ to the $i$-th vertex, and associate the variable $Y_{i,i+1}$ to the edge linking vertices $i$ and $i+1$.  In this case, it is easy to classify all possible region variables that remain in the symbol, and they are of the form
\begin{equation}
    X_{i,i+k}^{\pm\pm}:=\sum_{j=i}^{i+k} X_j\pm Y_{i-1,i}\pm Y_{i+k,i+k+1},
\end{equation}
where we take $Y_{0,1}=Y_{n,n+1}=0$, and we omit the $\pm$ index in the definition if the corresponding $Y$ variable vanishes. Importantly, the entries in the symbol can always be expressed as ratios of such variables. Therefore it is convenient to rescale the letters by the root variable $X_{1,n}$, and introduce a set of variables
\begin{equation}\label{eq:rescaled}
    x_{i,i+k}^{\pm\pm}=\frac{X_{i,i+k}^{\pm\pm}}{X_{1,n}}\,.
\end{equation}
As we mentioned before, these variables are not all independent, and in order to tackle the problem of integration of any symbol, it is necessary to reduce the problem to a minimal set of independent variables. Indeed, there are exactly $2n-1$ independent region variables, since $2n-1=|V_{P_n}|+|E_{P_n}|$, but only $2n-2$ independent rescaled variables \eqref{eq:rescaled}. The choice of an independent set is not unique. In the following, we  make a natural choice that will map the variables into those of the moduli space $\mathcal{M}_{0,2n+1}$. This can be accomplished by choosing the $2n-2$ variables to be
\begin{equation}
   y_j^{\pm}:= x_{1,j}^{\pm}=\frac{\sum_{i=1}^{j}X_i\pm Y_{j,j+1}}{X_{1,n}}, \quad j=1,2,\ldots,n-1\,.
\end{equation}
This minimal set has the remarkable property that it behaves exactly like simplicial coordinates: all other region variables that appear in the symbol are either the variables $y_{j}^\pm$ themselves, or differences of elements of this minimal set $y_{i}^\pm-y_{j}^{\pm}$, or of the form $1-y_{j}^\pm$. That is, the alphabet of the symbol organizes itself as the simplicial hyperplanes of the moduli space $\mathcal{M}_{0,2n+1}$ with $\{z_i\}_{i=1,2,\ldots,2n-2}=\{y_{j}^{\pm}\}_{j=1,2,\ldots,n-1}$. More precisely, the symbol letters are all of the form
\begin{equation}
    \{y_{j}^{\pm},1-y_{j}^{\pm},y_{j}^{\pm}-y_{k}^{\pm},y_{j}^{\pm}-y_{k}^{\mp}\}_{j\neq k\in\{1,\ldots, n-1\}}\,.
\end{equation}
Notice that not all hyperplanes are included in the alphabet. In particular, the $n-1$ combinations of the form $y_{i}^{+}-y_{i}^{-}$ are not present. This comes from the fact that $y_{i}^+-y_{i}^-=\frac{2Y_{i,i+1}}{X_{1,n}}$, where the numerator is a simple edge variable and not a region variable. 

To summarise, we have established that $\mathcal{X}$-coordinates of the cluster algebra $A_{2n-2}$ appear in the form of simplicial coordinates of the moduli space $\mathcal{M}_{0,2n+1}$ in the symbol of the path graph contribution to the wavefunction coefficient of a massless conformally coupled scalar in de Sitter space. Since the alphabet is the same, this observation extends to path graphs in FRW.

\subsection{Examples}
In the following, we provide two examples illustrating our statements, for the path graphs $P_2$ and $P_3$. 
\paragraph{Path graph $\mathbf{P_2}$}\,\\
Let us start with providing all details of our construction for the path graph $P_2$. First, there are three tubes that can be defined on $P_2$: 
\begin{equation}
    \{ {\begin{tikzpicture}[scale=0.6]
    \coordinate (A1) at (0,0);
    \coordinate (B1) at (1,0);
    \draw[thick] (A1) -- (B1);
        \draw[thick, black] (1/2,0) ellipse (0.75cm and 0.3cm);
    \fill[black] (A1) circle (2pt);
    \fill[black] (B1) circle (2pt);
    \end{tikzpicture}},{}, {}\}= \{X_{1,2},X_{1,1}^+,X_{2,2}^+\}\,,
\end{equation} 
and there are five regions 
\begin{equation}
    \{ {\begin{tikzpicture}[scale=0.6]
    \coordinate (A1) at (0,0);
    \coordinate (B1) at (1,0);
    \draw[thick] (A1) -- (B1);

    \draw[thick, black, fill=gray,opacity=0.5] (0,0) ellipse (0.15cm and 0.15cm);
            \draw[thick, black] (1/2,0) ellipse (0.75cm and 0.3cm);
    \fill[black] (A1) circle (2pt);
    \fill[black] (B1) circle (2pt);
\end{tikzpicture}}, {\begin{tikzpicture}[scale=0.6]
    \coordinate (A1) at (0,0);
    \coordinate (B1) at (1,0);
    \draw[thick] (A1) -- (B1);
        \draw[thick, black] (1/2,0) ellipse (0.75cm and 0.3cm);
    \draw[thick, black, fill=gray,opacity=0.5] (1,0) ellipse (0.15cm and 0.15cm);
    \fill[black] (A1) circle (2pt);
    \fill[black] (B1) circle (2pt);
\end{tikzpicture}},{\begin{tikzpicture}[scale=0.6]
    \coordinate (A1) at (0,0);
    \coordinate (B1) at (1,0);
            \draw[thick, red, fill=gray, opacity=0.5] (1/2,0) ellipse (0.75cm and 0.3cm);
    \draw[thick] (A1) -- (B1);
        \draw[thick, black] (1/2,0) ellipse (0.75cm and 0.3cm);
    \fill[black] (A1) circle (2pt);
    \fill[black] (B1) circle (2pt);
\end{tikzpicture}},{\begin{tikzpicture}[scale=0.6]
    \coordinate (A1) at (0,0);
    \coordinate (B1) at (1,0);
            \draw[thick, black, fill=gray, opacity=0.5] (1/2,0) ellipse (0.75cm and 0.3cm);
    \draw[thick] (A1) -- (B1);
        \draw[thick, black] (1/2,0) ellipse (0.75cm and 0.3cm);
            \draw[thick, black, fill=white] (1,0) ellipse (0.15cm and 0.15cm);
    \fill[black] (A1) circle (2pt);
    \fill[black] (B1) circle (2pt);
\end{tikzpicture}},{\begin{tikzpicture}[scale=0.6]
    \coordinate (A1) at (0,0);
    \coordinate (B1) at (1,0);
            \draw[thick, black, fill=gray, opacity=0.5] (1/2,0) ellipse (0.75cm and 0.3cm);
    \draw[thick] (A1) -- (B1);
        \draw[thick, black] (1/2,0) ellipse (0.75cm and 0.3cm);
            \draw[thick, black, fill=white] (0,0) ellipse (0.15cm and 0.15cm);
    \fill[black] (A1) circle (2pt);
    \fill[black] (B1) circle (2pt);
\end{tikzpicture}}\}=\{X_{1,1}^+,X_{2,2}^+,X_{1,2},X_{1,1}^-,X_{2,2}^-\}\,.
\end{equation}
Since the three tube variables are linearly independent in this case, the minimal set of region variables that we need to consider has two elements, as expected. We pick the following ratios of region variables
\begin{align}
    y_1^+=\frac{X_{1,1}^+}{X_{1,2}}\,, \qquad y_1^-=\frac{X_{{1,1}}^-}{X_{1,2}}\,.
\end{align}
Then, using these variables all rescaled letters for $P_2$ can be written as
\begin{equation}\label{eq: P2 alphabet}
  \left\{\frac{X_{1,1}^+}{X_{1,2}},\frac{X_{2,2}^+}{X_{1,2}},\frac{X_{1,1}^-}{X_{1,2}},\frac{X_{2,2}^-}{X_{1,2}}\right\} = \{y_1^+,1-y_1^-,y_1^-,1-y_1^+\}.
\end{equation}

\paragraph{Path graph $\mathbf{P_3}$}\, \\
Next, we consider the path graph $P_3$. In this case there are 6 tubes and 13 regions. The single linear relation between tube variables involves the only central tube 
\begin{equation}
X_{2,2}^{++}=X_{1,2}^++X_{2,3}^+-X_{1,3}\,,
\end{equation}
and it reduces the variables to 5 (that is 4 dimensionless ratios), as expected. We choose the following coordinates:
\begin{align*}
    y_1^+=\frac{X_{1,1}^+}{X_{1,3}}, \qquad y_1^-=\frac{X_{1,1}^-}{X_{1,3}}, \qquad y_{2}^+=\frac{X_{1,2}^+}{X_{1,3}}, \qquad y_{2}^-=\frac{X_{1,2}^-}{X_{1,3}}\,.
\end{align*}
In these variables, the rescaled alphabet for $P_3$ reads
\begin{align}\label{eq: P3 alphabet}
&\left\{\frac{X_{1,1}^+}{X_{1,3}},\frac{X_{1,1}^-}{X_{1,3}},\frac{X_{1,2}^+}{X_{1,3}},\frac{X_{1,2}^-}{X_{1,3}},\frac{X_{2,3}^-}{X_{1,3}},\frac{X_{2,3}^+}{X_{1,3}},\frac{X_{3,3}^-}{X_{1,3}},\frac{X_{3,3}^+}{X_{1,3}},\frac{X_{2,2}^{-+}}{X_{1,3}},\frac{X_{2,2}^{++}}{X_{1,3}},\frac{X_{2,2}^{--}}{X_{1,3}},\frac{X_{2,2}^{+-}}{X_{1,3}}\right\}\\
 &=   \{y_1^+,y_1^-,y_{2}^+,y_{2}^-, 1-y_1^+,1-y_1^-,1-y_{2}^+,1-y_{2}^-,y_{2}^+-y_1^+,y_{2}^+-y_1^-,y_{2}^--y_{1}^+,y_{2}^--y_{1}^-\}\,.\nonumber
\end{align}

\section{Applications}\label{sec:applications}
In this section, we leverage the connection we have found between the type-$A$ cluster algebras and the symbols for de Sitter wavefunction coefficients. There is a natural function space associated to each cluster algebra, that of cluster functions. Due to the relation discussed in section \ref{sec:cluster}, the treatment of type-$A$ cluster functions is reduced to that of polylogarithms in $\mathcal{M}_{0,n}$, which have been extensively studied in \cite{Brown:2009qja} and \cite{Bogner:2014mha}. In the following, after reviewing the definition of the relevant special functions, we present a minimal generating set of type-$A$ cluster polylogarithms and we subsequently describe how to integrate the symbol. See \cite{Parker:2015cia} for a similar application in $\mathcal{N}=4$ Yang-Mills with a thorough discussion of the same techniques.

\textit{Cluster functions} $f^{(w)}$ of transcendental weight $w$ satisfy the recursive property
\begin{equation}
 \text{d}f^{(w)}=\sum_{i}f_i^{(w-1)}\dlog{x_i}\,,
\end{equation}
where $x_i$ is a $\mathcal{X}$-coordinate of a cluster algebra. Due to the formula \eqref{eq: symbol recurrence}, this is equivalent to saying that the symbol of $f^{(w)}$ has depth $w$ and that it is constructed from cluster coordinates. It follows that the wavefunction coefficient $\psi_{P_n}$ is a weight-$n$ $A_{2n-2}$ function.
In simplicial coordinates, the problem is reduced to the integration of symbols in the moduli space $\mathcal{M}_{0,2n+1}$ \cite{Brown:2009qja,Bogner:2014mha}. It is the natural extension of the one-dimensional case, $\mathbb{P}_\mathbb{C}^1\backslash\{0,1,\infty\}$, where integrable symbols can always be written uniquely as combinations (sum and product) of the generalized polylogarithms $\text{Li}_{n_1,\dots, n_r}(z)$ \cite{Brown:2013qva}. In the multivariate case the space of functions extends to that of Goncharov polylogarithms, defined iteratively by
\begin{equation}
    G(a_1,\dots,a_k;z)=\int_0^z\frac{dt}{t-a_1}G(a_2,\dots,a_k;t)\,,
\end{equation}
where $G(;z):=1$. In particular, we have
\begin{align}
    G(\underbrace{0,\dots,0}_{k \text{ times}};z)=\frac{1}{k!}\log^k(z)\,, && G(\underbrace{1,\dots,1}_{k \text{ times}};z)=\frac{1}{k!}\log^k(1-z)\,,
\end{align}
and 
\begin{equation}
    G(\underbrace{0,\dots,0}_{(n_1-1)\text{ times}},1,\dots,1,\underbrace{0,\dots,0}_{(n_r-1)\text{ times}},1;z)=(-1)^r\,\Li_{n_r,\dots,n_1}(z)\,.
\end{equation}
Due to the shuffle relations among polylogarithms, the result cannot be written uniquely as a combination of the same functions, in contrast to the one-dimensional case. A basis can be given in terms of \textit{Lyndon words} \cite{RADFORD1979432, Brown:2009qja, Parker:2015cia, Weinzierl:2022eaz}.

A Lyndon word of length $k$ on an ordered set $S$ of size $n$, is a sequence $l$ of $k$ elements of $S$ such that $l$ is strictly smaller than any of its nontrivial cyclic permutations with respect to the lexicographic order in $S$. Type-$A$ cluster functions of any weight are generated by
\begin{equation}\label{eq: generating set}
    \bigcup_{k\in\mathbb{N}}\bigcup_{i=1}^n G(\text{Lyndon}_k(\{0,1,z_1,\dots,z_{i-1}\};z_i))\,,
\end{equation}
where $G(\text{Lyndon}_k(\{0,1,z_1,\dots,z_{i-1}\};z_i))$ denotes the set of Goncharov polylogarithms of weight $k$ whose first $k$ arguments form a Lyndon word on the set $\{0,1,z_1,\dots,z_{i-1}\}$. It follows from the discussion above that
the symbols of the de Sitter wavefunction coefficients associated to path graphs integrate to a unique combination of such functions. The symbol of each function of the basis can be found using \cite{Duhr:2012fh}
\begin{equation}\label{eq: symbols of Goncharovs}
\mathcal{S}(G(a_1,\dots,a_n;z))=\sum_{i=1}^n \mathcal{S}(G(a_1,\dots,\hat{a_i},\dots,a_n;z))\otimes \frac{a_i-a_{i-1}}{a_i-a_{i+1}}\,,
\end{equation} 
where $a_0=z$ and $a_{n+1}=0$, the hat indicates that the corresponding element is omitted, and
\begin{equation}\label{eq: shuffle of Goncharovs}
\mathcal{S}(G(a_1,\dots,a_r;z)G(a_{r+1},\dots,a_{r+s};z))=\sum_{\sigma\in\Sigma(r,s)} \mathcal{S}(G(a_{\sigma(1)},\dots, a_{\sigma(r+s)};z))\,,
\end{equation}
where $\Sigma(r,s)$ denotes the set of shuffles of two words of length $r$ and $s$.
Since \eqref{eq: generating set} is a generating set, any integrable symbol in simplicial coordinates can be expressed uniquely as a combination of symbols of the basis using linear algebra. This procedure gives the highest transcendental-weight part of the solution. In the following we apply this technique to integrate the symbols of $\psi_{P_2}$ and $\psi_{P_3}$.

\subsection{Path graph $P_2$}
The symbol of the wavefunction coefficient associated to the path graph $P_2$ in terms of its letters \eqref{eq: P2 alphabet} is
\begin{equation}
 \mathcal{S}(\psi_{\begin{tikzpicture}[scale=0.6]
    \coordinate (A) at (0,0);
    \coordinate (B) at (1/2,0);
    \draw[thick] (A) -- (B);
    \fill[black] (A) circle (2pt);
    \fill[black] (B) circle (2pt);
\end{tikzpicture}}^{\text{dS}})=y_1^+\otimes (1-y_1^+)+(1-y_1^-)\otimes y_1^--y_1^+\otimes(1-y_1^-)-(1-y_1^-)\otimes y_1^+\,.   
\end{equation}
This symbol will integrate to polylogarithms of weight up to 2. According to the generating set formula \eqref{eq: generating set}, to find the integrated answer we need to determine all Lyndon words of length $1$ and $2$ on $\{0,1\}$ and $\{0,1,y_1^+\}$\footnote{We could also change the order of simplicial coordinates. In this case, the final result is analogous.}. At weight 1 there are 5 functions
\begin{equation}
    \{G(0;y_1^+),G(1;y_1^+),G(0;y_1^-),G(1;y_1^-),G(y_1^+;y_1^-)\}\,.
\end{equation}
The functions of weight 2 are either products of those of weight 1 or pure weight 2 functions
\begin{equation}\label{eq: weight 2 basis}
    \{G(0,1;y_1^+),G(0,1;y_1^-),G(0,y_1^+;y_1^-),G(1,y_1^+;y_1^-)\}\,.
\end{equation}
Including the weight 0 constant, there is a total of 25 functions. The result presented in \cite{Hillman:2019wgh} written in our variables is 
\begin{equation}
 \psi^{\text{dS}}_{\begin{tikzpicture}[scale=0.6]
    \coordinate (A) at (0,0);
    \coordinate (B) at (1/2,0);
    \draw[thick] (A) -- (B);
    \fill[black] (A) circle (2pt);
    \fill[black] (B) circle (2pt);
\end{tikzpicture}}=-\Li_2(y_1^-)-\Li_2(1-y_1^+)-\Li_1(y_1^-)\Li_1(1-y_1^+)+\frac{\pi^2}{6}\,.
\end{equation}
Notice that $\Li_2(1-y_1^+)=-G(0,1;1-y_1^+)$ does not belong to the pure weight-2 function basis in \eqref{eq: weight 2 basis}. However, we have checked numerically that this result can be expanded in our basis as\footnote{For symbolic manipulation and the numerical evaluation of Goncharov polylogarithms in \texttt{Mathematica} we have used github.com/munuxi/Multiple-Polylogarithm by Zhenjie Li.} 
\begin{align}
    \psi^{\text{dS}}_{\begin{tikzpicture}[scale=0.6]
    \coordinate (A) at (0,0);
    \coordinate (B) at (1/2,0);
    \draw[thick] (A) -- (B);
    \fill[black] (A) circle (2pt);
    \fill[black] (B) circle (2pt);
\end{tikzpicture}}&=\Li_2(y_1^+)-\Li_2(y_1^-)+\log(1-y_1^+)\log(y_1^+)-\log(1-y_1^-)\log(y_1^+)\nonumber\\
    &=-G(0,1;y_1^+)+G(0,1;y_1^-)+G(1;y_1^+)G(0;y_1^+)-G(1;y_1^-)G(0;y_1^+)\,.
\end{align}

\subsection{Path graph $P_3$}
The symbol for $\psi_{P_3}$ in the alphabet \eqref{eq: P3 alphabet} is made up of 72 words of depth 3. Using Duval's algorithm \cite{DUVAL1988255}, we generate all Lyndon words of length 1, 2 and 3 on the sets $\{0,1\}$, $\{0,1,y_{1}^+\}$, $\{0,1,y_{1}^+,y_{1}^-\}$, $\{0,1,y_{1}^+,y_{1}^-,y_2^+\}$. The basis consists of 910 functions of weight-3, of which 70 are pure weight functions and 840 are products of the lower-weight ones. Applying \eqref{eq: symbols of Goncharovs} and \eqref{eq: shuffle of Goncharovs}, we find their symbols and their linear combination that gives the symbol of $P_3$. This finally gives
\begin{align}
\psi^{\text{dS}}_{\begin{tikzpicture}[scale=0.6]
    \coordinate (A) at (0,0);
    \coordinate (B) at (1/2,0);
        \coordinate (C) at (1,0);
    \draw[thick] (A) -- (B) -- (C);
    \fill[black] (A) circle (2pt);
    \fill[black] (B) circle (2pt);
        \fill[black] (C) circle (2pt);
\end{tikzpicture}}&=-G(0,1,y_1^-;y_2^-)+G(0,1,y_1^-;y^+_2)+G(0,1,y^+_1;y_2^-)-G(0,1,y^+_1;y^+_2)\nonumber\\
&-G(0,y_1^-,1;y_2^-)+G(0,y_1^-,1;y^+_2)+G(0,y^+_1,1;y_2^-)-G(0,y^+_1,1;y^+_2)\nonumber\\
&+G(1;y_2^-) G(0,1;y_1^-)-G(1;y^+_2)G(0,1;y_1^-)-G(y_1^-;y_2^-)G(0,1;y_1^-)+G(y_1^-;y^+_2)G(0,1;y_1^-)\nonumber\\
&+G(y_1^-;y_2^-)G(0,1;y_2^-)-G(y^+_1;y_2^-)G(0,1;y_2^-)-G(1;y_2^-)G(0,1;y^+_1)+G(1;y^+_2)G(0,1;y^+_1)\nonumber\\
&+G(y^+_1;y_2^-)G(0,1;y^+_1)-G(y^+_1;y^+_2)G(0,1;y^+_1)-G(y_1^-;y^+_2)G(0,1;y^+_2)+G(y^+_1;y^+_2)G(0,1;y^+_2)\nonumber\\
&+G(1;y_2^-)G(0,y_1^-;y^+_2)-G(1;y^+_2)G(0,y_1^-;y^+_2)-G(1;y_2^-)G(0,y^+_1;y^+_2)+G(1;y^+_2)G(0,y^+_1;y^+_2)\nonumber\\
&+G(0;y^+_1)G(1,y_1^-;y_2^-)-G(0;y^+_1)G(1,y_1^-;y^+_2)-G(0;y^+_1)G(1,y^+_1;y_2^-)+G(0;y^+_1)G(1,y^+_1;y^+_2)\nonumber\\
&+G(0;y^+_1)G(1;y_1^-)G(1;y^+_2)-G(0;y^+_1)G(1;y^+_1)G(1;y^+_2)+G(0;y^+_1)G(1;y_1^-) G(y_1^-;y_2^-)\nonumber\\
&-G(0;y^+_1)G(1;y_2^-)G(y_1^-;y_2^-)-G(0;y^+_1)G(1;y_1^-)G(y_1^-;y^+_2)+G(0;y^+_1) G(1;y_2^-)G(y_1^-;y^+_2)\nonumber\\
&-G(0;y^+_2) G(1;y_2^-)G(y_1^-;y^+_2)+G(0;y^+_2) G(1;y^+_2)G(y_1^-;y^+_2)+G(0;y^+_1) G(1;y_2^-)G(y^+_1;y_2^-)\nonumber\\
&-G(0;y^+_1) G(1;y^+_1)G(y^+_1;y_2^-)-G(0;y^+_1) G(1;y_2^-)G(y^+_1;y^+_2)+G(0;y^+_2) G(1;y_2^-)G(y^+_1;y^+_2)\nonumber\\
&+G(0;y^+_1) G(1;y^+_1)G(y^+_1;y^+_2)-G(0;y^+_2) G(1;y^+_2) G({y^+_1};y^+_2)+G({0};y_1^-) G({0};y^+_1)G({1};y_2^-)\nonumber\\
&+G({0};y^+_1) G({1};y_2^-)G({1};y^+_1)-G({0};y^+_1)G({1};y_1^-)G({1};y_2^-)-G({0};y_1^-) G({0};y^+_1)G({1};y^+_2)\nonumber\\
&-\frac{1}{2}G(0;y_1^-)^2 G({1};y_2^-)-\frac{1}{2}G({0};y^+_1)^2 G(1;y_2^-)\nonumber\\
&+\frac{1}{2}G({0};y_1^-)^2 G({1};y^+_2)+\frac{1}{2}G({0};y^+_1)^2 G({1};y^+_2)\nonumber\\
&+\text{lower-transcendental-weight terms}. 
\end{align}

\section{Conclusions}
\label{sec:concls}

In this paper we have presented a connection between cluster algebras and cosmological correlators for conformally coupled scalars in power-law FRW cosmologies. In particular, 
we have shown that the symbols of the wavefunction coefficients of path graphs $P_n$ are related to $A_{2n-2}$ cluster algebras. To this end we have used the relation between simplicial coordinates on the moduli space of genus-0 curves with marked points and $\mathcal{X}$-coordinates of type-$A$ cluster algebras. This connection provides alternative expressions for the integrated symbols in terms of cluster functions.

This work opens various possible directions of research. First, it would be interesting to extend this analysis to other graphs beyond $P_n$, such as more general tree-level graphs and graphs at loop level. Moreover, we have only presented some examples of integrated symbols for specific path graphs, and it is natural to wonder how to extend these results to generic $P_n$.
Finally, the natural question arises whether the notion of cluster adjacency is present also for cosmological correlators. As mentioned in the introduction, cluster algebras provide the singularities of scattering amplitudes and, through the adjacency property,  their interplay, dictating the sequence of such singularities in the symbol. The possibility of having cluster adjacency for cosmological correlators would provide powerful constraints on their possible analytic form. We leave this to future work.


\section{Acknowledgements}

This work was supported in part by the Deutsche Forschungsgemeinschaft (DFG, German Research Foundation) Projektnummer 508889767/FOR5582
``Modern Foundations of Scattering Amplitudes''.

\bibliographystyle{nb}
\bibliography{cluster_cosmo}

\end{document}